\documentclass[journal]{IEEEtran}

\ifCLASSINFOpdf
\else
   \usepackage[dvips]{graphicx}
\fi
\usepackage{url}
\usepackage{cite}
\usepackage{amsmath,amssymb,amsfonts}
\usepackage{algorithmic}
\usepackage{graphicx}
\usepackage{textcomp}
\usepackage{xcolor}
\usepackage{caption}
\usepackage{fixltx2e}
\usepackage{siunitx}
\usepackage{subfigure}
\usepackage{enumitem}
\usepackage{booktabs}
\usepackage{multirow}
\usepackage{multicol}
\usepackage{graphicx}

\begin{document}

\title{Tchebichef Transform Domain-based Deep Learning Architecture for Image Super-resolution}

\author{Ahlad Kumar and Harsh Vardhan Singh~\IEEEmembership{}

\thanks{Dr. Ahlad Kumar is an Assistant Professor  at Dhirubhai Ambani Institute of Information and Communication Technology (DA-IICT). (e-mail: ahlad\_kumar@daiict.ac.in)}
\thanks{Harsh Vardhan is currently pursuing his Master's degree in Information and Communication Technology from Dhirubhai Ambani Institute of Information and Communication Technology (DA-IICT), Gandhinagar, Gujarat (e-mail: harshsinghv@gmail.com )}
}

\maketitle

\begin{abstract}
The recent outbreak of COVID-19 has motivated researchers to contribute in the area of medical imaging using artificial intelligence and deep learning. 
Super-resolution (SR), in the past few years, has produced remarkable results using deep learning methods. The ability of deep learning methods to learn the non-linear mapping from low-resolution (LR) images to their corresponding high-resolution (HR) images leads to compelling results for SR in diverse areas of research. In this paper, we propose a deep learning based image super-resolution architecture in Tchebichef transform domain. This is achieved by integrating a transform layer into the proposed architecture through a customized Tchebichef convolutional layer ($TCL$). The role of $TCL$ is to convert the LR image from the spatial domain to the orthogonal transform domain using Tchebichef basis functions. The inversion of the aforementioned transformation is achieved using another layer known as the Inverse Tchebichef convolutional Layer ($ITCL$), which converts back the LR images from the transform domain to the spatial domain. It has been observed that using the Tchebichef transform domain for the task of SR takes the advantage of high and low-frequency representation of images that makes the task of super-resolution simplified. We, further, introduce transfer learning approach to enhance the quality of Covid based medical images. It is shown that our architecture enhances the quality of X-ray and CT images of COVID-19, providing a better image quality that helps in clinical diagnosis.  Experimental results obtained using the proposed Tchebichef transform domain super-resolution (TTDSR) architecture provides competitive results when compared with most of the deep learning methods employed using a fewer number of trainable parameters.

\end{abstract}

\begin{IEEEkeywords}
Image Super-Resolution (SR), Tchebichef moments, Deep Learning, Convolutional neural network, Covid-19.
\end{IEEEkeywords}

\IEEEpeerreviewmaketitle

\vspace{-2mm}
\section{Introduction}

\IEEEPARstart {T}{he} coronavirus disease (COVID-19) is a newly emerging viral disease that caused a worldwide pandemic. The  World Health Organization has listed it as the sixth international public health emergency. It has impacted around 170 countries and has almost taken the lives of 2 million people as of January 2021. COVID-19 diagnosis by X-ray and CT images is very popular due to the quick results and that with great accuracy. This paper proposes a deep learning based SR architecture to enhance the quality of COVID-19 medical images for clinical diagnosis.
 
Image Super-Resolution (SR) is one of the most famous and significant ill-posed problems since there can be multiple possible solutions existing for one single image. It refers to the process of obtaining high-resolution (HR) images from the corresponding low-resolution (LR) images. It's used in wide variety of applications such as in security, mobile cameras, medical imaging, etc. \cite{1} have attracted researchers over the past few years. SR problems are categorized into Single Image Super-resolution (SISR), and Multi-Image Super-resolution (MISR) \cite{2,3,4,5}. SISR aims to recover the HR image from a single LR image and the SISR based methods have an edge over MISR based methods, as they produce better perceptual quality of the images. The conventional SR based methods uses dictionary-based approaches, that consist of two dictionaries of HR and LR images or patches \cite{6,7,8,9}. These dictionaries are often learned with sparse-coding methods to get high-quality SR images. Some methods also use the dictionary features that need to be handcrafted \cite{11}. Recent advancements in deep learning methods have shown the state of the art results in SR over the different dataset of images \cite{12}. The earliest deep learning method was SRCNN\cite{13} and later its improved version was reported in \cite{17}. Recursive networks\cite{18} and residual learning\cite{19} is also incorporated in the deep learning architectures to boost the training process.

We propose a new orthogonal domain based deep learning architecture for image SR. It uses Tchebichef moment to transform spatial domain to orthogonal domain and then finds the difference between the Tchebichef coefficients of HR and LR image pairs. It has been observed that HR-LR image pairs have huge differences in the coefficient values at higher frequencies and small to negligible differences in the lower frequencies. This key observation plays a central role in developing the proposed architecture. The key aspects of this paper are summarized as follows
\begin{itemize}
  \item A deep neural network to solve the SR problem in an orthogonal transform domain is introduced. The architecture includes both the forward and inverse mapping, hence provides a complete pipeline for image SR.
  \item The proposed architecture exploits the Tchebichef kernels to generate the representation of images in the transform domain. Two custom convolutional layers are designed; one for converting the image to transform domain ($TCL$) and the other, for doing inverse transformation ($ITCL$). $TCL$ layer is kept fixed and non-trainable, while the $ITCL$ as trainable to get the optimized reconstruction kernels, which are used in $ITCL$ layer for converting from transform domain to spatial domain.
 
  \item The proposed architecture consist of high and low-frequency paths. The high frequency path employs Inception-Resnet based structure with local residual connection to boost the training process. The low-frequency path employs a simple convolutional neural network (CNN) architecture.
  
  \item To handle the artifacts that occur during the reconstruction phase after ITCL, additional convolutional layers are employed to process the spatial domain images that results in the enhanced SR images. 
  

\end{itemize}

\section{Related Work}
\subsection{Single Image Super Resolution (SISR)}
The SISR  poses a highly challenging problem that is not so easy to solve due to the ill-posed nature, and hence, SISR involves multiple solutions for one LR image or patch. SISR can mainly be divided into three categories: learning, interpolation and reconstruction-based methods.

Learning-based SISR methods \cite{7,55} are computationally fast and have good performance, specifically, sparse-coding has shown compelling results. The method involves two dictionaries of images, separately for LR and HR patches. LR image or patch are represented in terms of their corresponding sparse code from the LR dictionary. Similarly, an HR image or patch is generated by the same sparse code, but it is applied to the HR dictionary. 
Interpolation SISR methods involve techniques, like bicubic interpolation \cite{21} and Lanczos resampling \cite{22}, although they are speedy and straightforward, but lack satisfactory accuracy. Reconstruction based methods \cite{23,24,25,26} are time-consuming, SISR relies on advantages from prior knowledge but these methods highly degrade the SR image quality when the scaling factor is increased.


\subsection{Deep Learning Advancements in Image Super Resolution}
Earlier due to lack of computational capabilities and not so advanced architectures, the conventional SR methods have produced average results. Recent advancements both in computational architectures and effective research using deep neural networks, have produced state-of-the-art results in the SR domain. Dong $et \; al.$ \cite{13} introduced a CNN network for the SR problem, known as SRCNN. It involved three layers of architecture that outperformed the previous methods based on sparse-coding. SRCNN introduced the non-linear mapping function between the LR and HR image. Image patches are fed to the CNN to generate the feature representation followed by more convolutional layers to produce the higher representation of images. Although SRCNN produced great results, but continuous advancements in the SR domain led to newer methods and architectures.

Since then fully connected neural networks have also shown considerable improvements in SR by combining several ideas from different architectures; such as sparse-coding, residual learning, etc. Prior information based deep neural networks has shown promising results. FSRNet \cite{32} used to generate the human face SR images. Wang $et \; al.$ \cite{33} focused on using the structural feature priors for effective recovery of detailed texture features in an image.

A huge improvement in SR image quality was observed using Generative Adversarial Networks (GANs) \cite{34}. Johnson $et\; al.$ \cite{35} focused on creating the photo-realistic image with high perceptual quality and focused less on maintaining the pixel-wise difference between the images. GANs generate visually better images but training a GAN is a complex and time-consuming process making it unsuitable for many practical applications.

Li $et\; al.$ \cite{67} introduced the combined architecture using image transform domain as well as CNN, and converted the input image/patch into the Fourier transform domain. The discrete Fourier transform (DFT) coefficients, thus obtained are passed to a CNN architecture. As the convolution of the kernels and image in the spatial domain is equivalent to the multiplication of kernels and images in Fourier domain, the claim is that the CNN architecture performs the element-wise multiplication to speed up the training process.  Experimentally, it was observed that the performance was similar to previous works but not at par with state-of-the-art methods. Similar work is reported in \cite{20} which uses a combination of transform domain and CNN, where the images are transformed using discrete cosine transform (DCT) and incorporated the pre and post-processing of DCT and inverse discrete cosine transform (IDCT) coefficients within the CNN, using custom convolutional layer.

This paper exploits the image transform domain using Tchebichef basis functions that are  integrated as kernels with the CNN network. A custom convolutional layers ($TCL$) for transforming the image into orthogonal domain and the reverse using inverse transform ($ITCL$) is proposed. The transform domain representations of an image is utilized to learn the high-frequency details that are lost during the degradation using bicubic interpolation. The inverse transformation layer ($ITCL$) is kept trainable and optimizable, hence the kernels used in $ITCL$ optimize with training process. The optimized kernels obtained after training process provides the best possible reconstruction of the images from its corresponding transform domain. Moreover, the reconstruction artifacts like softness and haziness are removed by the help of additional layers, incorporated in the proposed architecture. The main objective of this paper is to performs SR effectively by preserving the visual attributes of an image.
\section{Tchebichef Moments}
\subsection{Computation of Tchebichef Moments }
In this section, a brief review about the definition of Tchebichef moment \cite{36} is discussed. 
They have been used recently in many pattern recognition  \cite{61,62,63} and denoising applications \cite{64,65}. Its robust feature representation capabilities allows to reconstruct images with promising results. The Tchebichef moment of order $(m+n)$ for an image with intensity function $g(x,y)$ is given as \cite{36}
\begin{equation}
T_{n,m}(u)=\sum_{x=0}^{N-1}\sum_{y=0}^{N-1}\widetilde{t}_{m}(x;N)\widetilde{t}_{n}(y;N)g(x,y)
\label{tu}
\end{equation}
with $n,m = 0,1,2, \ldots, N-1$. The image $g(x,y)$, being of size $ N\times N$,  $\widetilde{t}_{n}(x;N)$ and $\widetilde{t}_{m}(y;N)$ are the normalized Tchebichef  polynomials,  given by 
\begin{equation}
\widetilde{t}_{m}(x;N)=\frac{t_{m}(x;N)}{\rho(m,N)}
\end{equation}
and 
\begin{equation}
\widetilde{t}_{n}(y;N)=\frac{t_{n}(y;N)}{\rho(n,N)}
\end{equation}
with $\rho(m,N)$=(2$m$)!$N+m \choose 2m+1$, $\rho(n,N)$=(2$n$)!$N+n \choose 2n+1$ and $t_{n}(x;N)$ is the $\textit{N}$-th order $\textit{N}$-point Tchebichef polynomial defined as 
\begin{equation}
t_{n}(x;N)=n!\sum\limits_{k=0}^{n}(-1)^{n-k}\binom{N-1-k}{n-k} \binom{n+k}{n}\binom{x}{k}\nonumber
\end{equation} 
To simplify the notation, $t_{n}(x)$ is used to represent $t_{n}(x;N)$. Here, $t_{n}(x)$ is the ortho-normal version of Tchebichef polynomials and it can be calculated using recurrence relation as \cite{36} 
\begin{equation}
\begin{aligned}
\tilde{t}_{n}(x) &=\alpha_{1}(2 x+1-N) \tilde{t}_{n-1}(x)+\alpha_{2} \tilde{t}_{n-2}(x) \\
\end{aligned}
\end{equation}

where
\begin{equation}
\begin{aligned}
\alpha_{1} &=\frac{1}{n} \sqrt{\frac{4 n^{2}-1}{N^{2}-n^{2}}} \\
\alpha_{2} &=\frac{1-n}{n} \sqrt{\frac{2 n+1}{2 n-3}} \sqrt{\frac{N^{2}-(n-1)^{2}}{N^{2}-n^{2}}}
\end{aligned}
\end{equation}

The initial conditions for the above recurrence relationship is given as
\begin{equation}
\begin{aligned}
\tilde{t}_{0}(x) & =1 / \sqrt{N} \\
\tilde{t}_{1}(x) & =(2 x+1-N) \sqrt{\frac{3}{N\left(N^{2}-1\right)}}
\end{aligned}
\end{equation}

Image can be reconstructed back from Tchebichef moments using the inverse Tchebichef transformation as given by
\begin{equation}
g(x,y)=\sum_{n=0}^{N-1}\sum_{m=0}^{N-1}\widetilde{t}_{m}(x;N)\widetilde{t}_{n}(y;N)T_{n,m}(u)
\label{nj}
\end{equation}

\subsection{Matrix Form}
Tchebichef moment in (\ref{tu}) can also be implemented in matrix format. The set of Tchebichef moments upto order $(m+n)$ in matrix form is given as
\begin{equation}
    \textbf{T} = \textbf{P} \textbf{G}\textbf{Q}^{T}
\label{c1p}
\end{equation}
where $\textbf{G}$ is a square image matrix. $\textbf{P}$ and $\textbf{Q}$ are Tchebichef polynomials in matrix form upto orders of $p$ and $q$, respectively given as
\begin{equation}
\textbf{P} = \begin{bmatrix} 
    \tilde{t}\textsubscript{0}(0) & \dots  & \tilde{t}\textsubscript{0}(N-1) \\
    \vdots & \ddots & \vdots\\
    \tilde{t}\textsubscript{p}(0) & \dots  & \tilde{t}\textsubscript{p}(N-1)
    \end{bmatrix}
\label{}
\end{equation}

\begin{equation}
\textbf{Q} = \begin{bmatrix} 
    \tilde{t}\textsubscript{0}(0) & \dots  & \tilde{t}\textsubscript{0}(N-1) \\
    \vdots & \ddots & \vdots\\
    \tilde{t}\textsubscript{q}(0) & \dots  & \tilde{t}\textsubscript{q}(N-1)
    \end{bmatrix}
\label{}
\end{equation}
Similarly, the inverse transformation given in (\ref{nj}) can be represented in matrix form as
\begin{equation}
    \textbf{G} = \textbf{P}^{T} \textbf{T}\textbf{Q}
\label{c1p}
\end{equation}

\subsection{Basis functions of Tchebchief Moment}
\begin{figure}[h!]
  \centering
  \includegraphics[width=2in,height=2in]{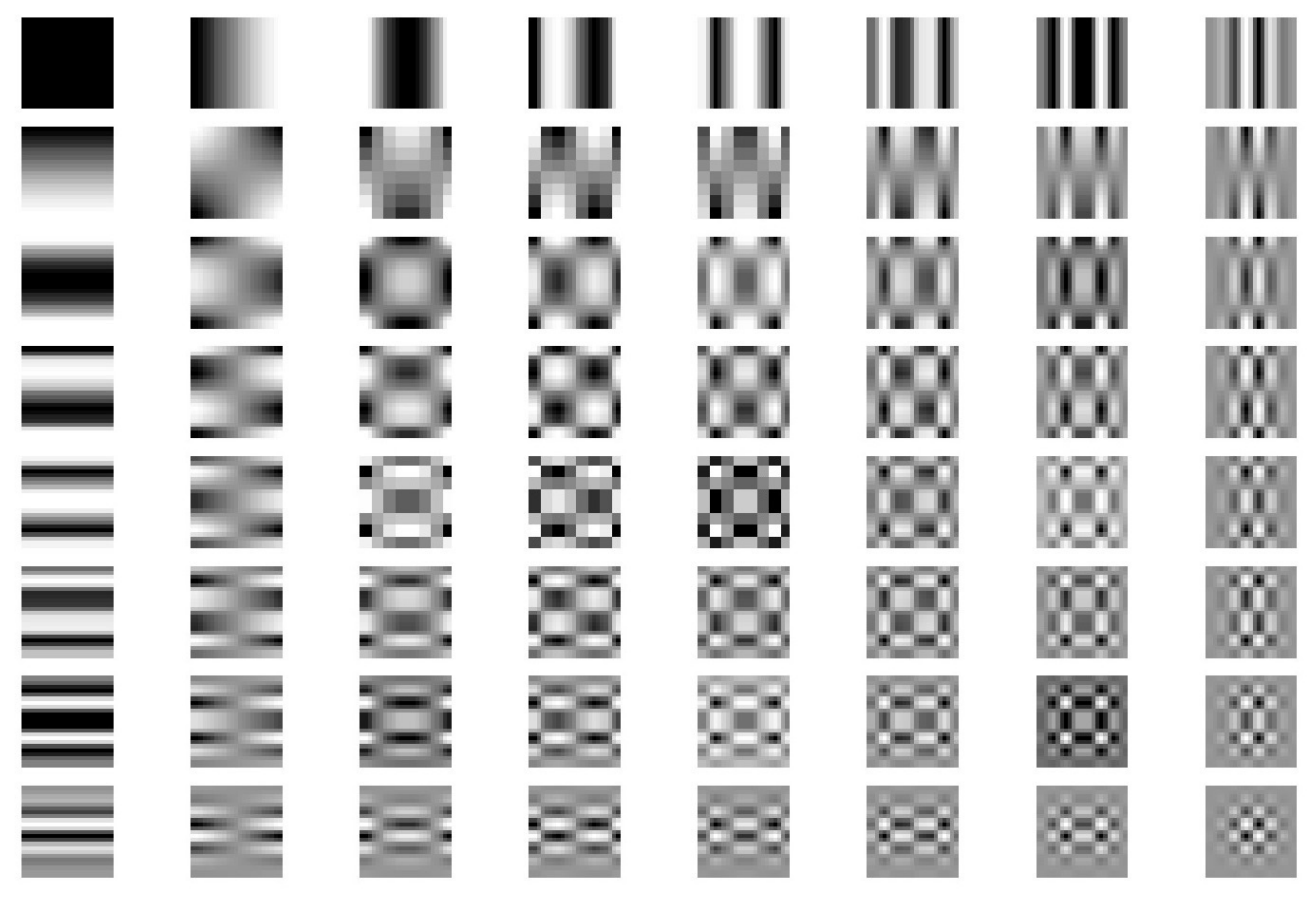}
  \caption{Basis functions of Tchebichef moment}
  \label{fx1}
\end{figure}

Tchebichef moments of an image can be interpreted as the projection of the image on the basis (kernel) functions, $\mathbf{w}_{pq}$ which is given as
\begin{equation}
\mathbf{w}_{\mathbf{p q}} =\left[\tilde{\mathbf{t}}_{\mathbf{p}}\right]^{T} \tilde{\mathbf{t}}_{\mathbf{q}}
\end{equation}
where
\begin{equation}
\begin{aligned}
\tilde{\mathbf{t}}_{\mathbf{p}} &=\left[\tilde{t}_{p}(0) \;\tilde{t}_{p}(1) \ldots\tilde{t}_{p}(N-1)\right] \\
\tilde{\mathbf{t}}_{\mathbf{q}} &=\left[\tilde{t}_{q}(0)\; \tilde{t}_{q}(1) \ldots \tilde{t}_{q}(N-1)\right] \\
\end{aligned}
\end{equation}
The complete set of $\mathbf{w}_{pq}$ basis functions is shown in Fig. \ref{fx1}. Tchebichef moments can also be viewed as correlation between the basis functions and image $\textbf{G}$. A high value is recorded, if there is a strong similarity between the content of the image and basis function, and vice versa.

\subsection{Basis Ordering and its Significance} \label{g1}
\begin{figure}[h!]
  \centering
  \includegraphics[width=2in,height=2in]{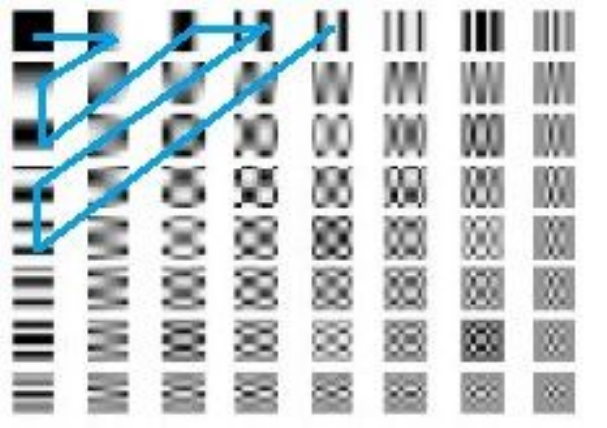}
  \caption{ Zig-zag re-ordering of Tchebichef basis functions}
\label{fx2}
\end{figure}

In the proposed architecture, Tchebichef basis functions are used as filters and are re-arranged in a zig-zag order as shown in Fig. \ref{fx2} This zig zag reordering is inspired from the JPEG compression procedure \cite{40}. 

\begin{figure*}[!ht]
\centering
\subfigure[]{\includegraphics[width=2.35in]{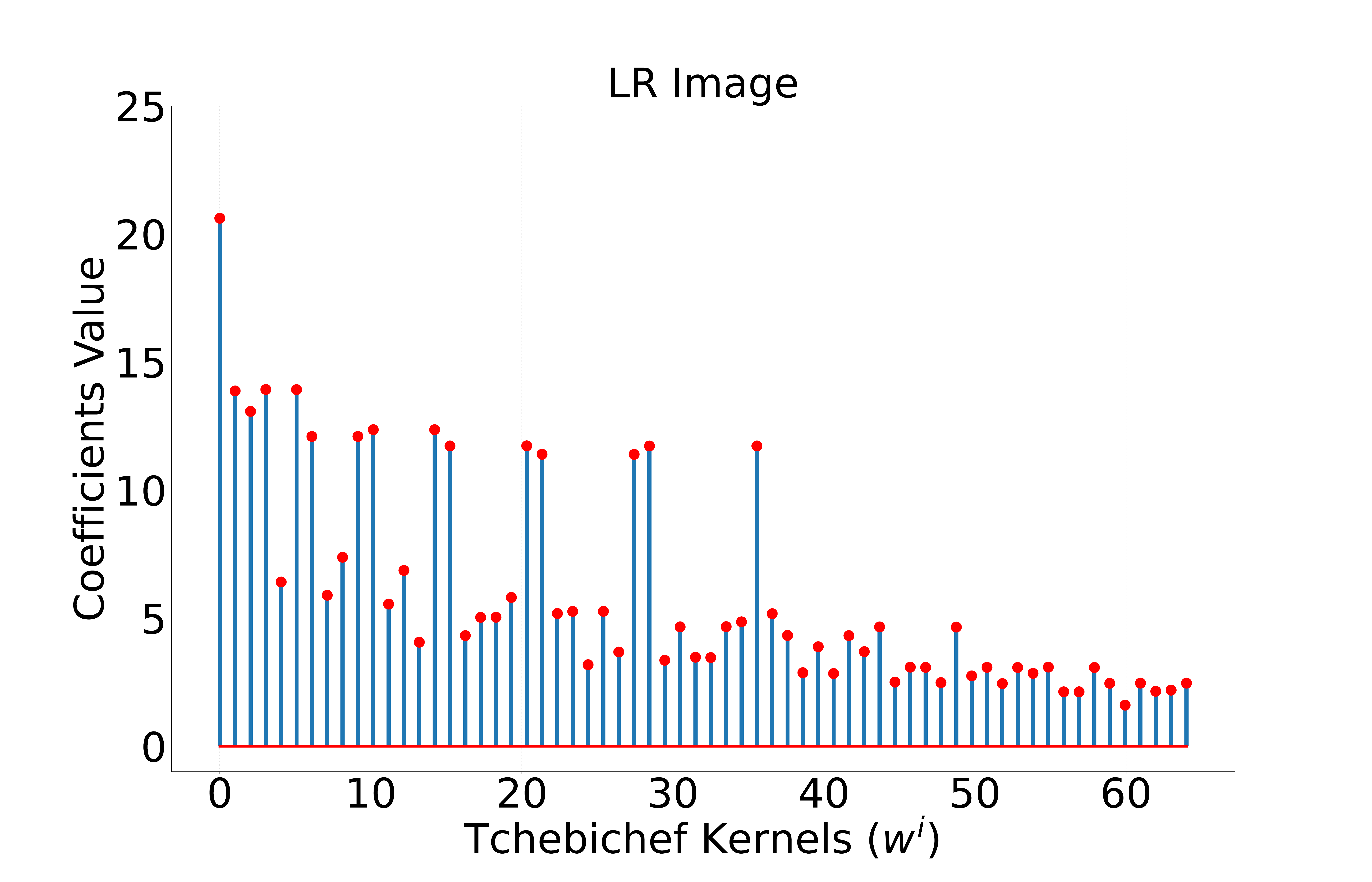}}
\subfigure[]{\includegraphics[width=2.35in]{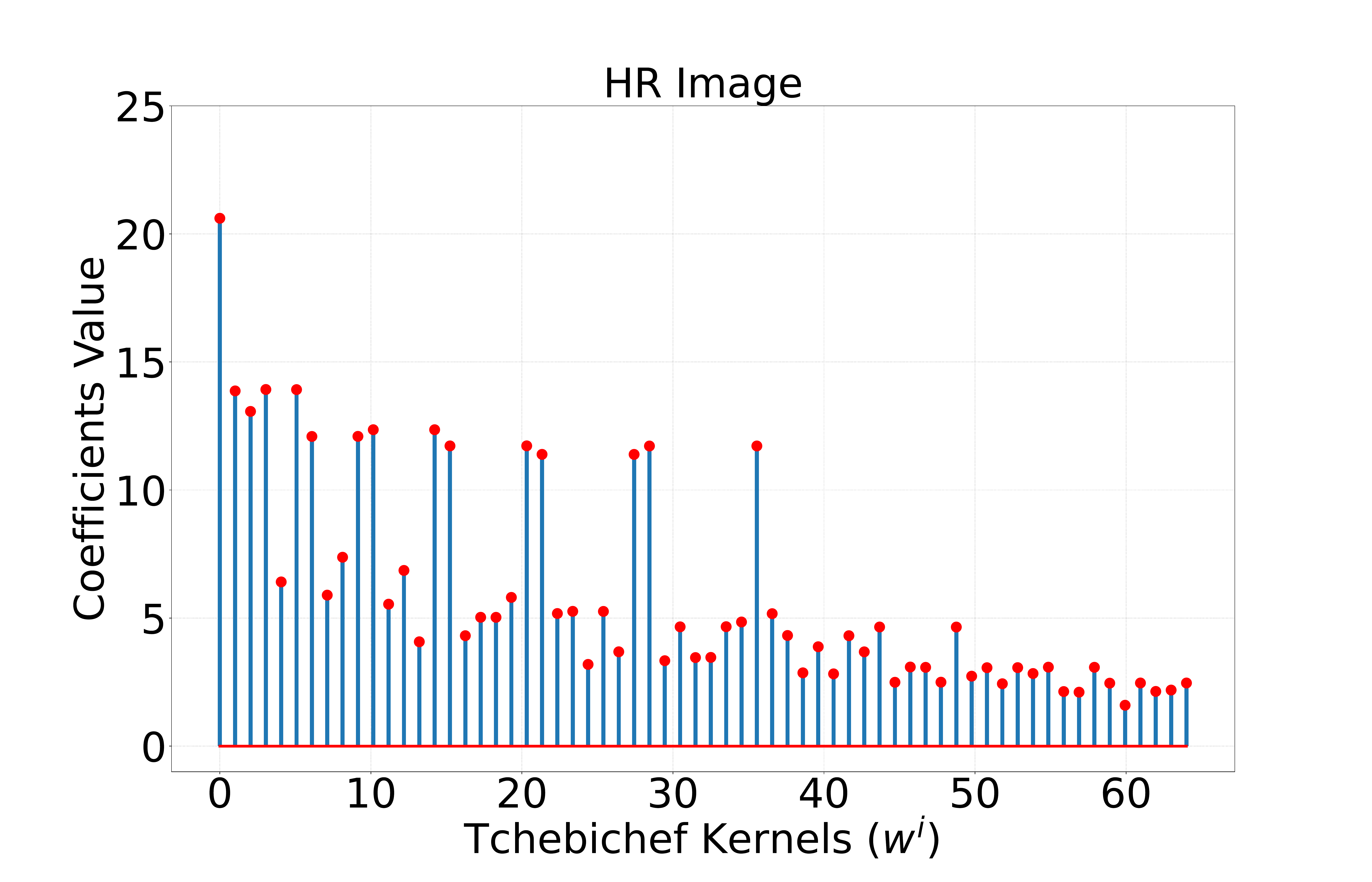}}
\subfigure[]{\includegraphics[width=2.35in,height=1.52in]{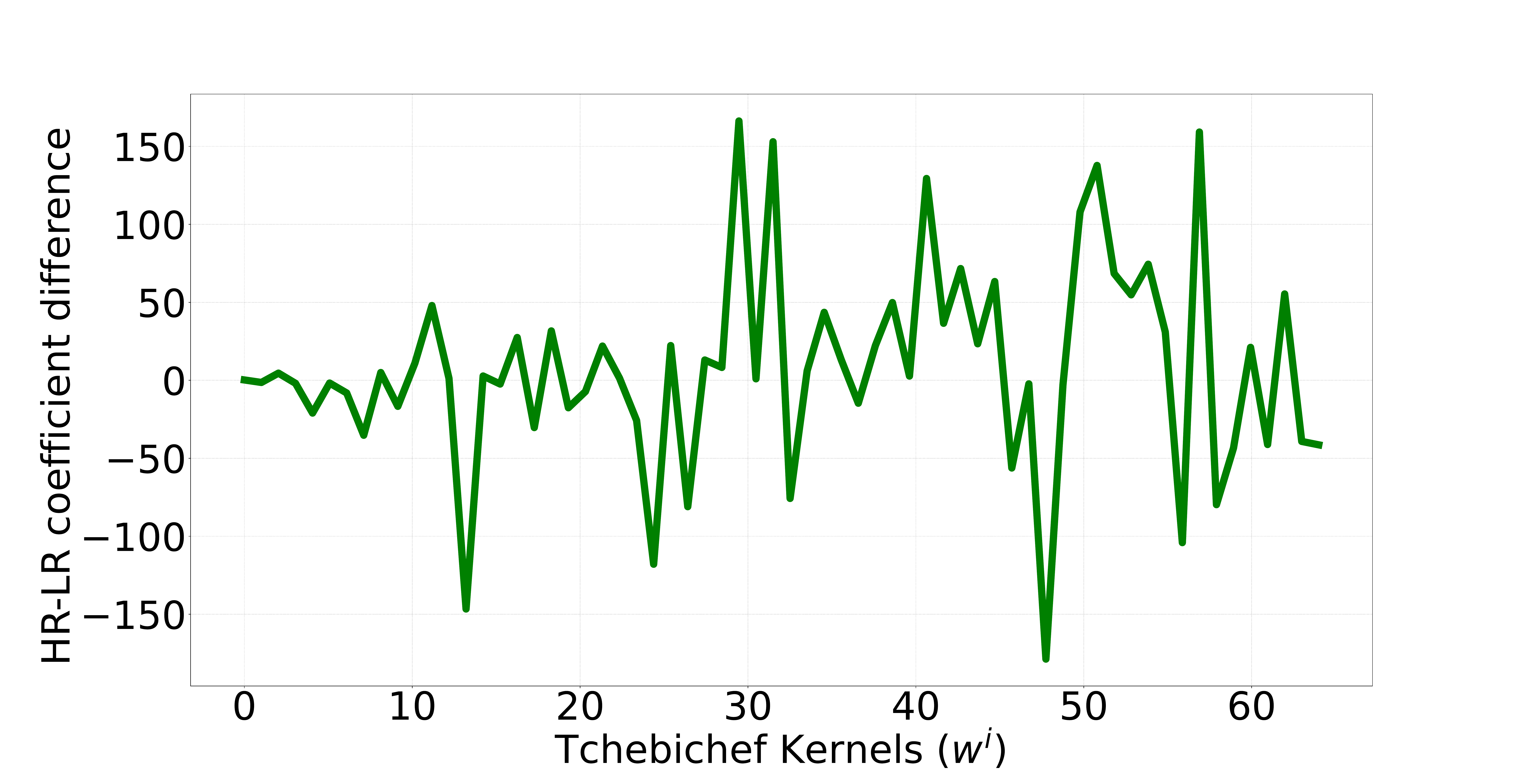}}\\
\subfigure[]{\includegraphics[width=2.35in]{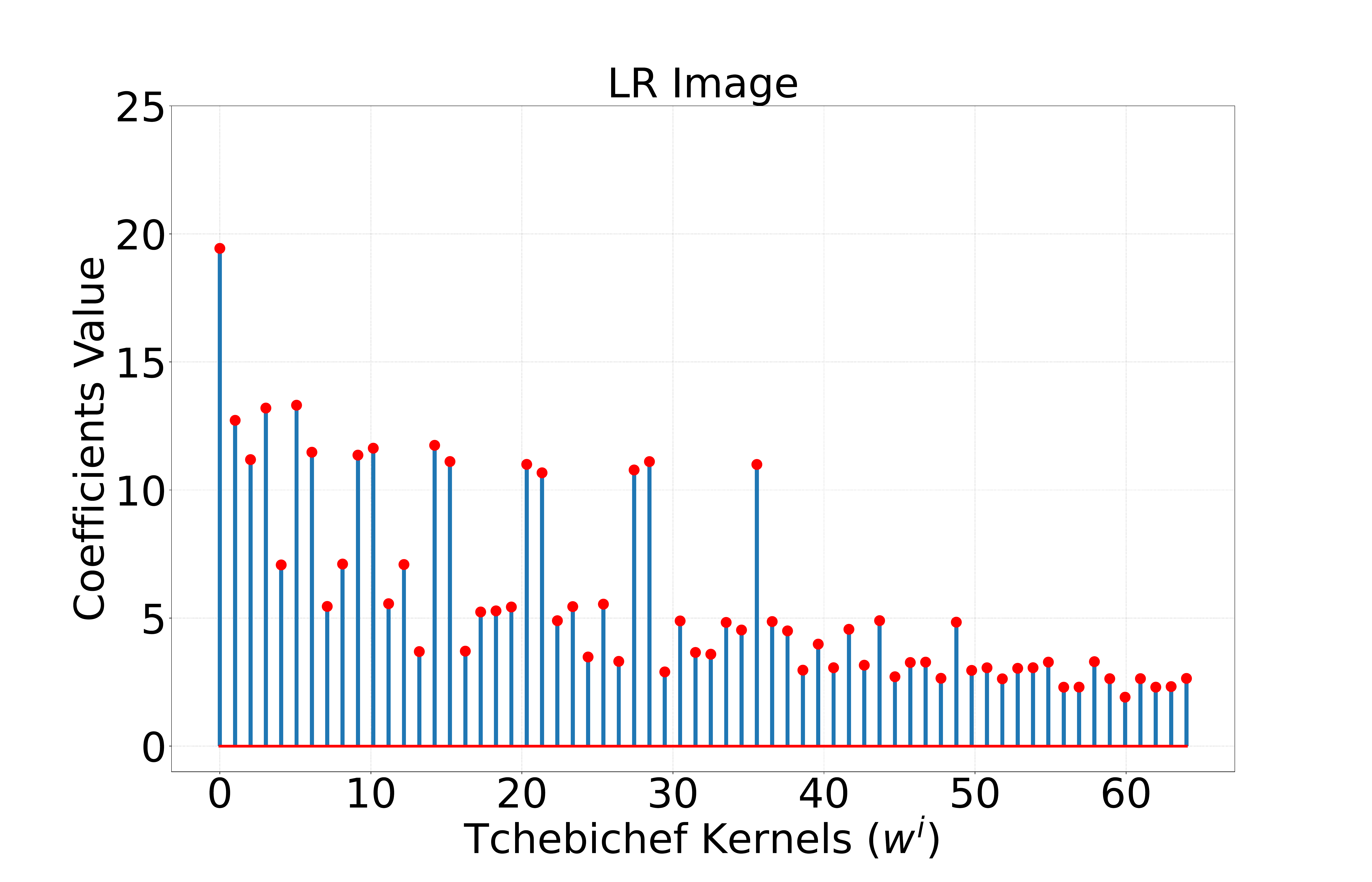}}
\subfigure[]{\includegraphics[width=2.35in]{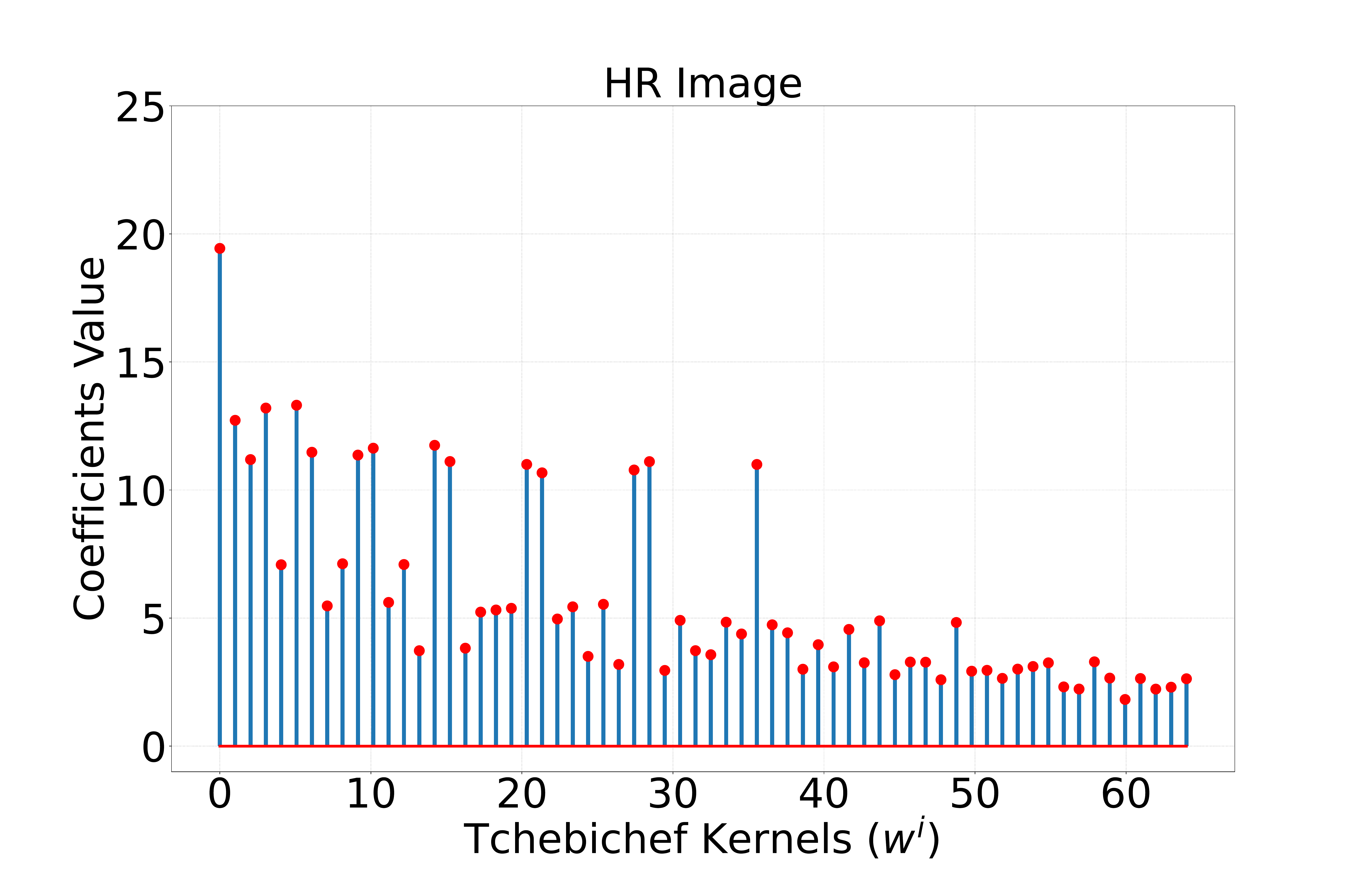}}
\subfigure[]{\includegraphics[width=2.35in]{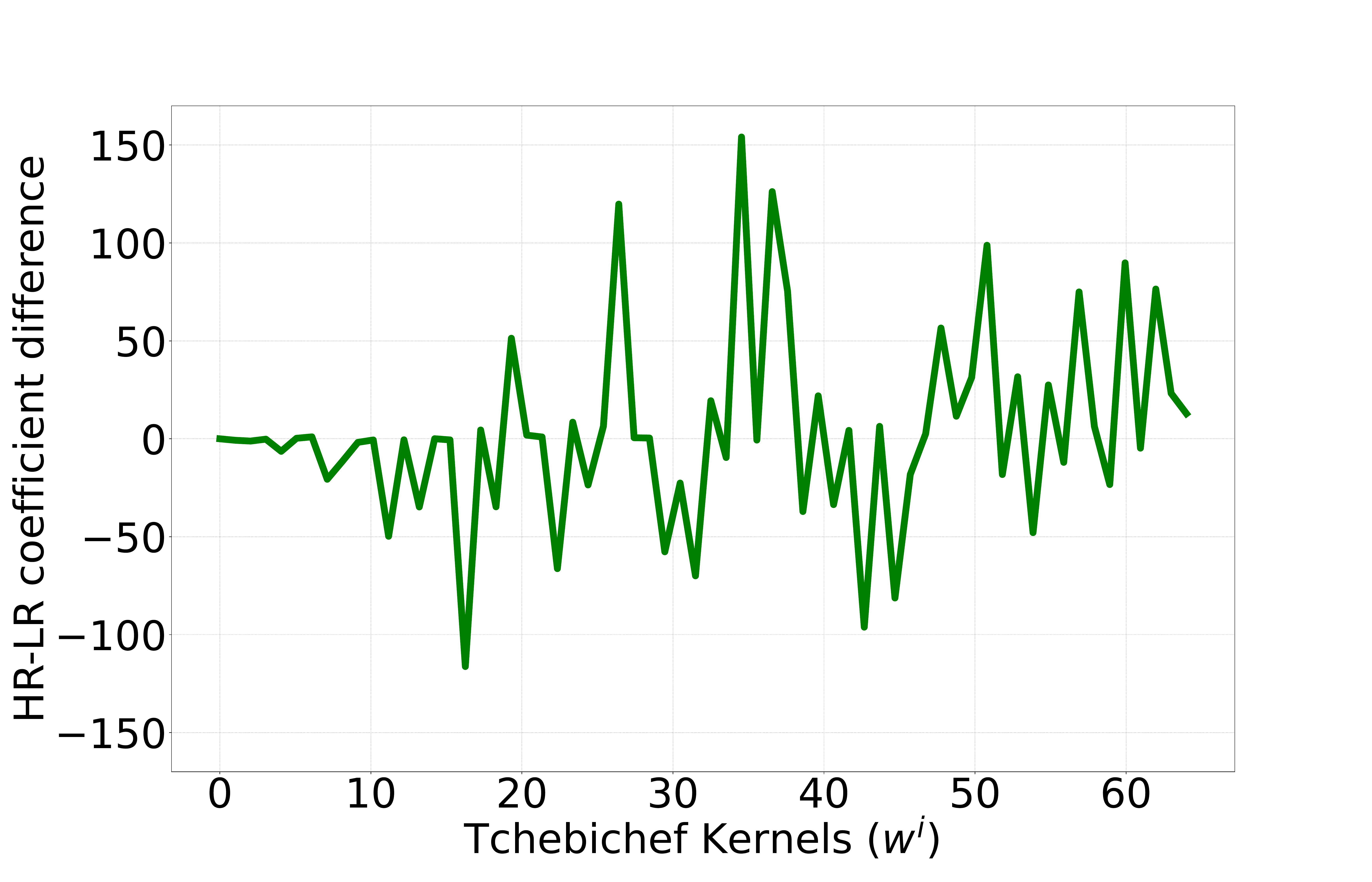}}
 \caption{Top row: Tchebichef coefficients for (a) LR (b) HR (c) Difference in the coefficient of HR and LR for Medical image.
Bottom row: Tchebichef coefficients for (d) LR  (e) HR (f) Difference in the coefficient of HR and LR for natural image}
\label{fx3}
\end{figure*} 
Zig-Zag ordering of the basis functions allows to exploit the transform domain efficiently. We represent the 64 zig-zag reordered basis functions as $\mathbf{w}^{i}$ where $i$ = 0 to 63. It is observed that this particular re-ordering of basis functions results in an increase frequency pattern (complexity) in the basis functions, i.e, as the index $i$ increases, the frequency content increase from low to high. 

The average value of the coefficients generated by the convolution of the Tchebichef kernels with HR and LR images respectively, is shown in Fig. \ref{fx3}. Here, Figs. \ref{fx3}(a)-(b) show the coefficients of a LR and HR version of the medical image, while Fig. \ref{fx3}(c) shows the differences between the HR and LR image coefficients. Please note that the values obtained in Fig. \ref{fx3}(c) is scaled for proper visualization. It can be observed from this figure that with the increase in kernels complexity, there is a substantial loss of the coefficients in the high frequency when compared to lower frequency region. In the Tchebichef domain, the problem of SR becomes recovering the high frequency Tchebichef coefficients of the HR image from their corresponding LR images. This observation is incorporated in the proposed architecture discussed next. Similar analysis is carried out on natural image and the results are shown in Figs. \ref{fx3}(d)-(f).

\begin{figure*}[!ht]
    \includegraphics[width=173mm,height=95mm]{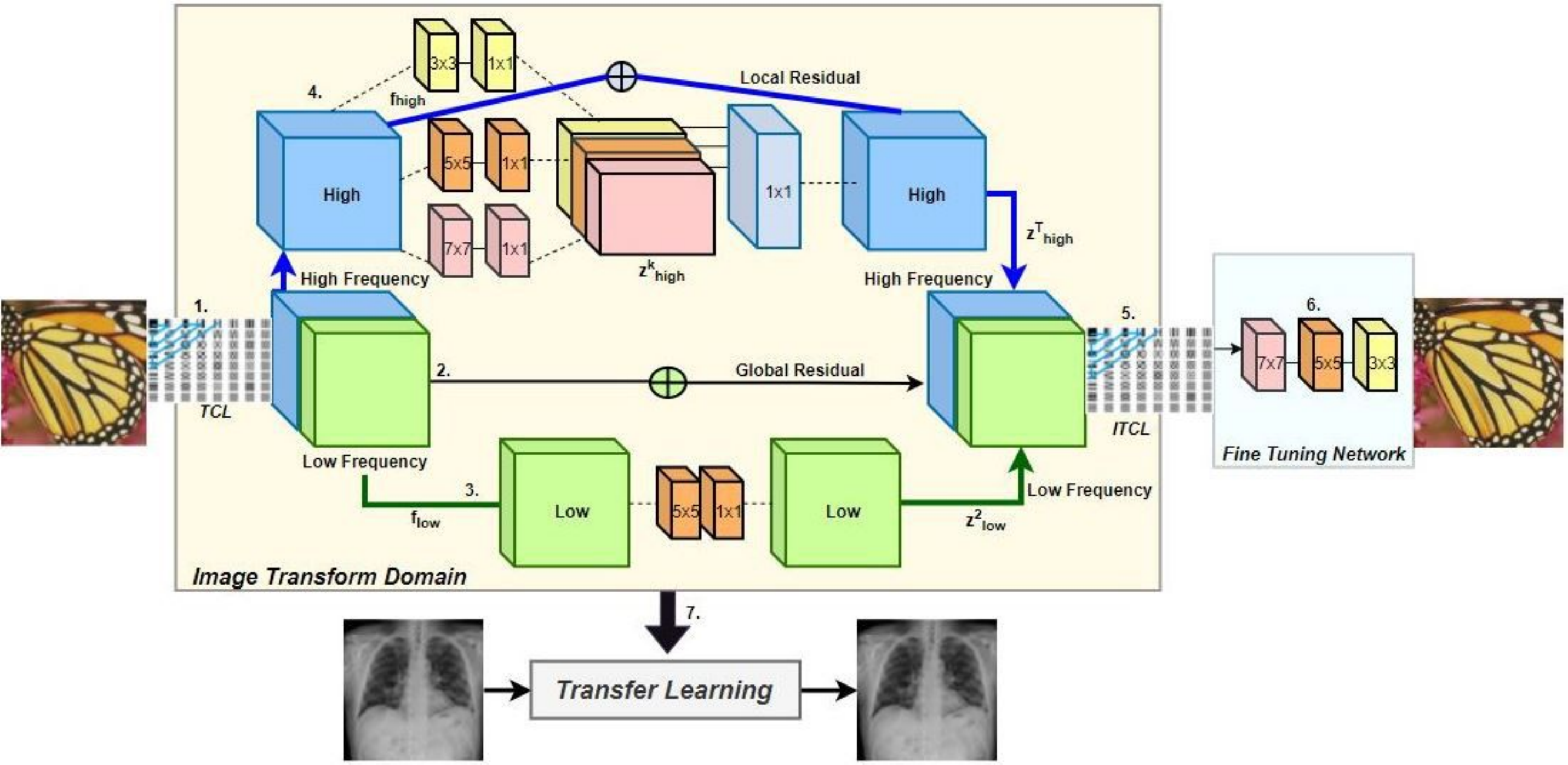}
\caption{The proposed TTDSR network architecture. Please refer to the number markings used in explanation of architecture. Blue and green color are used for the high and low frequency details respectively. Yellow color background shows entire processing in the transform domain and light blue is used for the spatial domain.}
\label{fx4}
\end{figure*}

\section{Proposed Tchebichef Transform Domain Super Resolution (TTDSR)}
In this section, a detail description of the proposed architecture, shown in Fig. \ref{fx4} for SR is discussed. The architecture consist of the following blocks: (1) Tchebichef convolutional layer ($TCL$) (2) Frequency cube (3) Non-linear mapping for low frequency. (4) Inception-residual connection for high frequency (5) Inverse Tchbeichef transformation layer ($ITCL$). 
\subsection{Network Structure}
\subsubsection{Tchebichef convolutional layer ($TCL$)}
This block transforms images from spatial domain to Tchebichef moment domain, and has basis function $\mathbf{w}_{i}$ as the kernels. There are 64 such kernels of size 8$\times$8 arranged in a zig-zag manner for increasing complexity with an increase in index $i$ of the kernels. The detail about this is discussed later in Sec. \ref{g1}.

The transformation from spatial to Tchebichef moment domain works as follows: $TCL$ layer creates a 64 feature maps $\mathbf{f}_{i}$ for the entire image by performing convolution using $\mathbf{w}_{i}$ with image  \textbf{G} as given in (\ref{c1}). Here, $\circledast$ represent the convolution operation and is performed using the stride $S = 1$ and $same$ padding in order to preserve the dimension of the image.

\begin{equation}
\mathbf{f}_{i}=\mathbf{w}_{i} \circledast \textbf{G}, \quad \forall i \in\{1, \ldots, 64\}\\
\label{c1}
\end{equation}
Kernels of the $TCL$ layer are kept fixed and non-trainable during the training phase as the primary role of this layer is to convert images into the transform domain. 
\subsubsection{Frequency Cube}
\begin{figure}[h!]
  \centering
  \includegraphics[width=200pt,height=150pt]{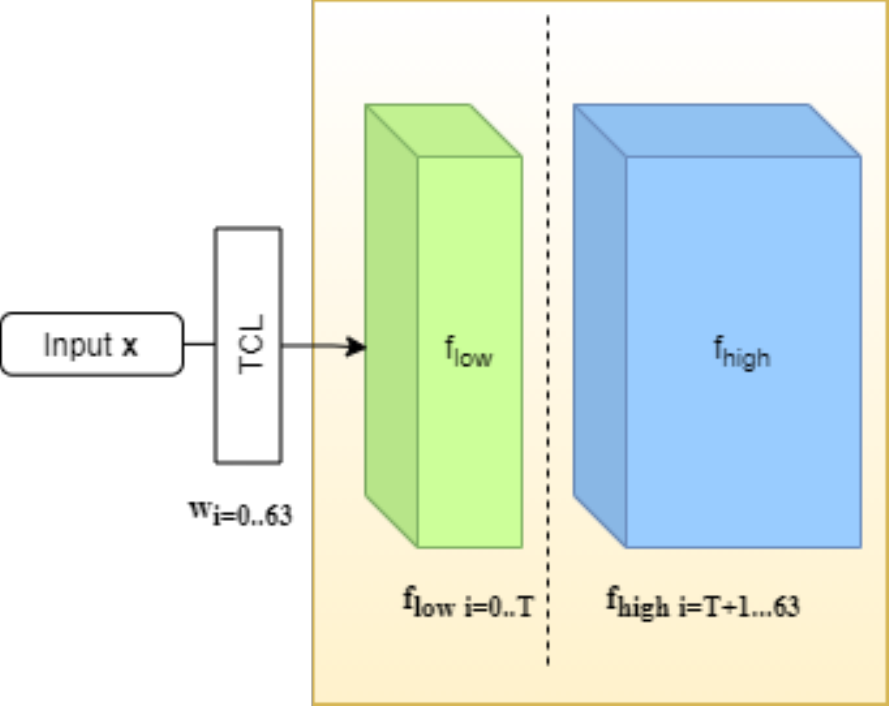}
  \captionof{figure}{The cube is partitioned at a split point T to generate the high frequency and low frequency cube.}
\label{fx5}
\end{figure}

Frequency domain feature maps $\textbf{f}_{i = 0, \ldots, 63}$, obtained from (\ref{c1}), are used to form a cube (see label 2 marked in Fig. \ref{fx4}). This cube is re-organized version of Tchebichef coefficients, calculated for the whole image and is ordered in increasing frequency content (complexity). Based on the detail discussion carried out for Fig. (\ref{fx3}) in Sect. \ref{g1}, it has been observed that there is a substantial loss of the coefficients in the high frequency region compared to lower frequency region. Due to this reason, partition of the frequency cube is done into two parts with a particular split point $T$. The low and high frequency maps are defined as $\textbf{f}_{low} = \textbf{f}_{i=1,\ldots, T}$ and $\textbf{f}_{high} = \textbf{f}_{i=T+1,\ldots, 63}$, respectively. Fig. \ref{fx5} shows the details about this partitioning process. The calculation of the split point $T$ is experimental and its optimal value is obtained as 5. The discussion about its optimal value is carried out in experimental section.

The proposed architecture process the partitioned cubes $\textbf{f}_{low}$  and  $\textbf{f}_{high}$ separately. It can be observed from Figs. \ref{fx3}(c) and (f), that there is a higher amount of coefficient loss in high-frequency region and thereby the high frequency block $\textbf{f}_{high}$ requires more robust and complex mapping to recover the HR image from LR image. On the other hand, the coefficient loss in low-frequency region is not so significant, but does play an important role in image quality. Hence, processing the low frequency block $\textbf{f}_{low}$ is done, using simpler non-linear convolutional mapping to recover the image details. Next, we discuss the simple and complex deep learning architectures for $\textbf{f}_{low}$ an $\textbf{f}_{high}$.
\subsubsection{Architecture for $\textbf{f}_{low}$} \label{v1}
The mapping  of low-frequency coefficients of LR image to corresponding low frequency coefficients of HR image is accomplished via CNN network consisting of two convolutional layers (see green arrow in Fig. \ref{fx4}). The first layer is a 5$\times$5 followed by 1$\times$1 convolutional layer. Leaky rectified linear unit (ReLU) is used as the activation function in both the layers. The non-linear mapping is given as 
\begin{equation}
\begin{aligned}
\textbf{z}^{[0]}_{low} & = \mathbf {f}_{low} \\
\textbf{z}^{[k]}_{low} & =\max\left(\mathbf {z}^{[k-1]}_{low} \circledast \mathbf{W}_{1}^{[k]}+\mathbf{B}_{1}^{[k]}, \;\mathbf{\alpha}\;\textbf{f}_{low} \right)
\quad k \in\{1,2\}
\label{cv1}
\end{aligned}
\end{equation}
where $k$ represents the index to the two convolutional layers, $\textbf{z}^{[k]}_{low}$ is the output of $k^{\text{th}}$  layer, $\mathbf{W}_{1}^{[k]}$ and $\textbf{B}_{1}^{[k]}$ are the  weights and biases of the $k^{\text{th}}$  layer, $\alpha$ is the leaky ReLU parameter having value of 0.1. The non-linear mapping of (\ref{cv1}) recovers the information loss in the lower frequency spectra of the image  

\subsubsection{Architecture for $\textbf{f}_{high}$ } \label{v2}
To recover the information loss in high frequency spectra of the image, a non-linear mapping is implemented using the deep learning architecture inspired from the inception network \cite{41} (see green arrow in Fig. \ref{fx4}). The high frequency feature maps $f_{high}$ of LR image is split into three convolutional paths; each of which are using different kernel sizes, namely, 3$\times$3, 5$\times$5, and 7$\times$7. Larger kernel sizes are used in gathering the global information, while the smaller kernel size gather information that is distributed more locally in the feature map. This allows the model to take advantage of the multi-level feature extraction. Finally, concatenating the features obtained from all the levels is done followed by a 1$\times$1 convolution, which serves two purposes. First it creates a linear projection of a stack of feature maps, and secondly, it reduces the depth of the network. The non-linear mapping for the process described above is given as 
\begin{equation}
\textbf{z}^{[k]}_{high}=\max \left(\textbf{f}_{high} \circledast \mathbf{W}_{2}^{[k]}+\mathbf{B}_{2}^{[k]},\;\mathbf{\alpha\; \textbf{f}_{high}}\right), 
\quad k \in\{1,2,3\}
\end{equation}

\begin{equation}
\textbf{z}^{T}_{high}= \sum_{k=1}^{3} \textbf{z}^{[k]}_{high}
\end{equation}
Here, $\textbf{z}^{T}_{high}$ is the combination of all the feature maps obtained via three parallel paths denoted by $k$.

\vspace{10pt}
\subsubsection{Inverse Tchebichef Transformation Layer (ITCL)}
This layer is required to transform image from Tchebichef moment domain to spatial domain. It takes  input which is obtained by combining both, the low and high frequency cubes, $\textbf{z}^{[2]}_{low}$ and $\textbf{z}^{T}_{high}$, respectively.  The output of this layer reconstruct the image $\widehat{\textbf{y}}$ in spatial domain and is given as
\begin{equation}
\hat{\mathbf{y}}=\sum_{i=1}^{64} \mathbf{w}_{i} \circledast \left(\textbf{z}^{[2]}_{low} + \textbf{z}^{T}_{high}\right)_{i}
\label{kl}
\end{equation}
Here, the weights of the Tchebichef kernel $w_{i}$ are trainable, so that during the training process the kernels adapt to the data and provides efficient inverse transformation.

\subsubsection{Fine-Tuning Network}
The reconstructed image $\widehat{\textbf{y}}$ obtained using (\ref{kl}) is further processed through a small fine tuning network shown in Fig. \ref{fx4}, which consist of three convolutional layers. The main purpose of introducing this additional network is to get rid of minor artifacts from the image $\widehat{\textbf{y}}$.

\section{Experimental Work}

\subsection{Training Details}
This section discuss the training details of the 
proposed TTDSR architecture. In order to learn the end to end mapping function $\textbf{F}$ for SR task, optimized values of the network parameters $\theta \in (\mathbf{W}_{1}^{[k]}, \mathbf{B}_{1}^{[k]}, \mathbf{W}_{2}^{[k]}, \mathbf{B}_{2}^{[k]}$) are required. These parameters can be obtained by minimizing the loss between the network generated reconstructed SR image $\textbf{F}(\mathbf{Y}_{i},\mathbf{\theta})$,  and the high-resolution ground truth image $\textbf{X}$. Given the batch of high-resolution images $\textbf{X}_{i}$ and the corresponding low-resolution images $\textbf{Y}_{i}$, the loss function is given as

\begin{equation}
L(\theta)=\frac{1}{M} \sum_{i=1}^{M}\left\|\textbf{F}\left(\mathbf{Y}_{i} ; \theta\right)-\mathbf{X}_{i}\right\|^{2} + \mathbf \lambda \sum_{j=1}^{l} W_{j}^{2}\\
\end{equation}
where $M$ is the total number of training images, $\lambda$ is the regularization parameter and $l$ is the total number of kernels used in the architectures discussed in Sections \ref{v1} and \ref{v2}. The loss is minimized using the Adaptive moment estimation i.e Adam \cite{43} optimizer with standard back-propagation \cite{44}, that computes adaptive learning rates of each parameter. The training phase of the network involves Adam as optimizer with default parameters as: $\beta_{1} = 0.9$ (the first-order moments), $\beta_{2} = 0.999$ (the second-order moments), $\epsilon = 1\mathrm{e}{-7}$ (a small constant for numerical stability). Learning rate $\eta$ is initialised as $1\mathrm{e}{-3}$. The filter weights for each layer in the network are initialized with Glorot-uniform that draws the samples from the uniform distribution. It is observed that without using regularization, the network becomes highly un-stable, hence, L2 regularization of $\lambda = 0.01 $  is applied on the network weights to penalize the weights.
The $TCL$ is kept as non-trainable, while the $ITCL$ is kept as trainable to get the best optimized Tchebichef kernels.
There are 14 convolutional layers in the TTDSR architecture
leading to total number of parameters as 94k, out of which 90k are the trainable parameters and the remaining are fixed parameters which are used in the $TCL$ layer. The network is trained for 100 epochs with batch size of 64. Training and testing phase are conducted on NVIDIA GeForce RTX 2080 Ti GPU, with the TensorFlow as support package.

\subsection{Datasets} \label{kai}
There are several widely used dataset for image SR. The combination of images from T91 \cite{45} and DIV2K \cite{12} dataset have been used to create a combined training dataset. From a robust model, data augmentation technique is used, where the images are augmented using three methods, i.e,

\begin{enumerate}[label={\alph*.}]
  \item
  The images are rotated using 45\textdegree, 90\textdegree, 135\textdegree, 180\textdegree, 225\textdegree.
  \item 
  Horizontal and vertical flips of the images are done.
  \item 
  Scaling of images is performed by factor of 0.6, 0.7, 0.8 and 0.9 respectively.
\end{enumerate}

These augmented images are the variations in the HR image, which are then down-sampled by a factor of $\eta$. The down-sampled images are scaled up using the bi-cubic interpolation of same factor $\eta$ to form the degraded LR images for training. Training images are first converted from $RGB$ to $YCbCr$ format. Inspired from \cite{9,19,46}, luminance ($Y$) channel is used as the input to the architecture while $Cb$ and $Cr$ channels are directly up-scaled, using the bi-cubic interpolation of LR images. Finally, combining the up-scaled $Cb$ and $Cr$ channels, with the predicted luminance $(Y)$, channel from the architecture is used to generate the SR image which is then converted back into $RGB$ format. As the proposed architecture is trained on single channel, i.e., $Y$ in $YCbCr$ domain, this allow for the flexibly of performing transfer learning on Covid-19 medical images, which are gray-scale images. SR results for the same can be analyzed in the experimental work.

\begin{figure}[h!]
  \centering
  \includegraphics[width=250pt,height=170pt]{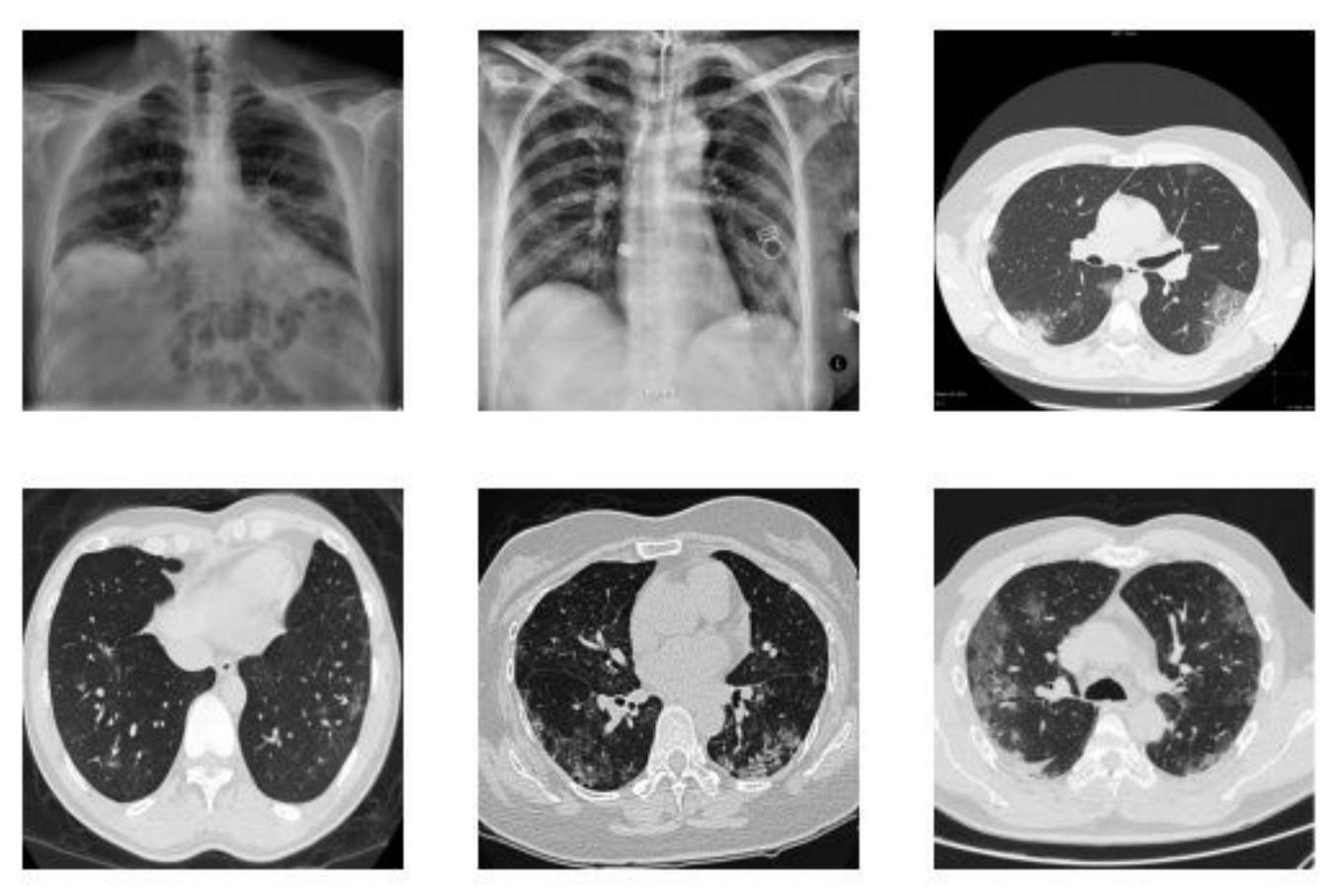}
  \captionof{figure}{Sample images from COVID-19 dataset which contains both X-ray and CT images.}
\label{fx6}
\end{figure}

During the testing phase, several standard datasets like Set5 \cite{47}, Set14 \cite{48}, BSDS100 \cite{49} and Urban100 \cite{50} have been utilized to evaluate the performance of the proposed architecture. 
The metrics used for image quality assessment are PSNR and SSIM \cite{51}. Few published methods work with larger datasets like DIV2K \cite{12}, ImageNet\cite{52} and MS-COCO\cite{53}. However, our choice of datasets for comparison has been used to keep it consistent with the majority of the methods that make use of it. The above datasets used for testing mainly consist of natural images. As pointed out earlier, transfer learning is carried out on the proposed architecture, which is now tested on medical images. For this, COVID-19 image database which contains a set of images collected by Cohen $et\;al.$ \cite{54} are used. The dataset contains of chest X-ray and computed tomography (CT) images. The images are mainly in gray-scale format and is a collection of anterior-posterior view of chest X-rays. The dataset is continuously updated and it is worth mentioning that the resolution of images varies from image to image. A sample of these images can be found in Fig. \ref{fx6}. 

\subsection{Comparative Analysis}
In this section, the performance of the proposed architecture with the other existing methods is discussed.
For this, several standard datasets (Sect. \ref{kai}) are being used for its evaluation. Here, following methods are used for comparison with our architecture
\begin{enumerate}[label={\alph*.}]
  \item
  ScSR \cite{9} : Sparse coding based SR method, constructs LR-HR image patch dictionary.
  \item
  A+ \cite{57} : Adjusted anchored neighbourhood regression for fast super resolution is the updated and modified version of \cite{60}.
  \item
  SelfEx \cite{50} : Self similarity based method that measures the similarity within the images.
  \item
  SCN \cite{58} : Sparse prior method implemented with the help of CNN.
  \item
  SRCNN \cite{13} : Earliest deep learning method for image SR based on CNN architecture.
  \item
  FSRCNN \cite{17} : An advanced and modified version of SRCNN with deeper architecture and transpose convolution approach.
  \item
  VBPS \cite{59} : Recent method for image SR that exploits the inherent self-similarities found in images.
\end{enumerate}

\begin{table*}[!htbp]
\centering
\renewcommand{\arraystretch}{1.0}
\captionsetup{justification=centering, labelsep=newline,font=footnotesize,labelfont=normalsize}
\caption{PSNR COMPARISON FOR VARIOUS SR METHODS. BLUE  REPRESENTS FIRST AND RED REPRESENTS SECOND BEST.}
\begin{tabular}{ccccccccccc}
 \specialrule{1.5pt}{1pt}{1pt} \\
 \multirow{2}{*}{Dataset} & Scale & Bicubic & ScSR & A+ & SelfEx & SCN & SRCNN & FSRCNN & VBPS &  TTDSR\\
 & & & \cite{9} &\cite{57} &\cite{50} &\cite{58} &\cite{13} &\cite{17} & \cite{59} & (Proposed) \\
 \hline \\
 \multirow{3}{*}{Set5} & 2x & 33.64 & 35.78 & 36.55 & 36.50 & 36.58 & 36.66 & \color{red}36.94 & \color{blue}36.97 & 35.35 \\
    & 3x & 30.39 & 31.34 & 32.58 & 32.62 & 32.61 & 32.75 & 33.06 & \color{red}33.23 & \color{blue}\textbf{33.50}\\
    & 4x & 28.42 & 29.07 & 30.27 & 30.32 & 30.41 & 30.48 & 30.55  & \color{red}31.19 & \color{blue}\textbf{32.41}\\
 \hline \\
  \multirow{3}{*}{Set14} & 2x & 30.22 & 31.64 & 32.29 & 32.24 & 32.35 & 32.42 & 32.54 & \color{red}32.83 & \color{blue}\textbf{33.47} \\
    & 3x & 27.53 & 28.19 & 29.13 & 29.16 & 29.16 & 29.28 & 29.37 & \color{red}29.61 & \color{blue}\textbf{32.26}\\
    & 4x & 25.99 & 26.40 & 27.33 & 27.40 & 27.39 & 27.40 & 27.50 & \color{red}27.86 & \color{blue}\textbf{31.55}\\
 \hline \\
  \multirow{3}{*}{BSDS100} & 2x & 29.55 & 30.77 & 31.21 & 31.18 & 31.26 & 31.36 & \color{red}31.66 & - &\color{blue}\textbf{33.15}\\
    & 3x & 27.21 & 27.72 & 28.18 & 28.30 & 28.58 & 28.20 & \color{red}28.52 & - & \color{blue}\textbf{32.03} \\
    & 4x & 25.96 & 26.61 & 26.82 & 26.84 & 26.88 & 26.84 & \color{red}26.92 & - & \color{blue}\textbf{31.68}\\
 \hline \\
 \multirow{3}{*}{Urban100} & 2x & 26.66 & 28.26 & 29.20 & 29.54 & 29.52 & 29.50 & 29.87 & \color{red}30.36 & \color{blue}\textbf{32.32}\\
    & 3x & 24.46 & 25.34 & 26.03 & 25.69 & 25.56 & 26.24 & 26.35 & \color{red}26.94 & \color{blue}\textbf{31.69}\\
    & 4x & 23.14 & 24.02 & 24.32 & 24.78 & 25.13 & 24.52 & 24.61 & \color{red}25.12 & \color{blue}\textbf{31.41}\\
 \specialrule{1.5pt}{1pt}{1pt} \\
 
\end{tabular}
\label{tab:1}
\end{table*}

\begin{figure*}[!htbp]
\centering
\begin{tabular}{cc}
    \includegraphics[width=155mm,height=45mm]{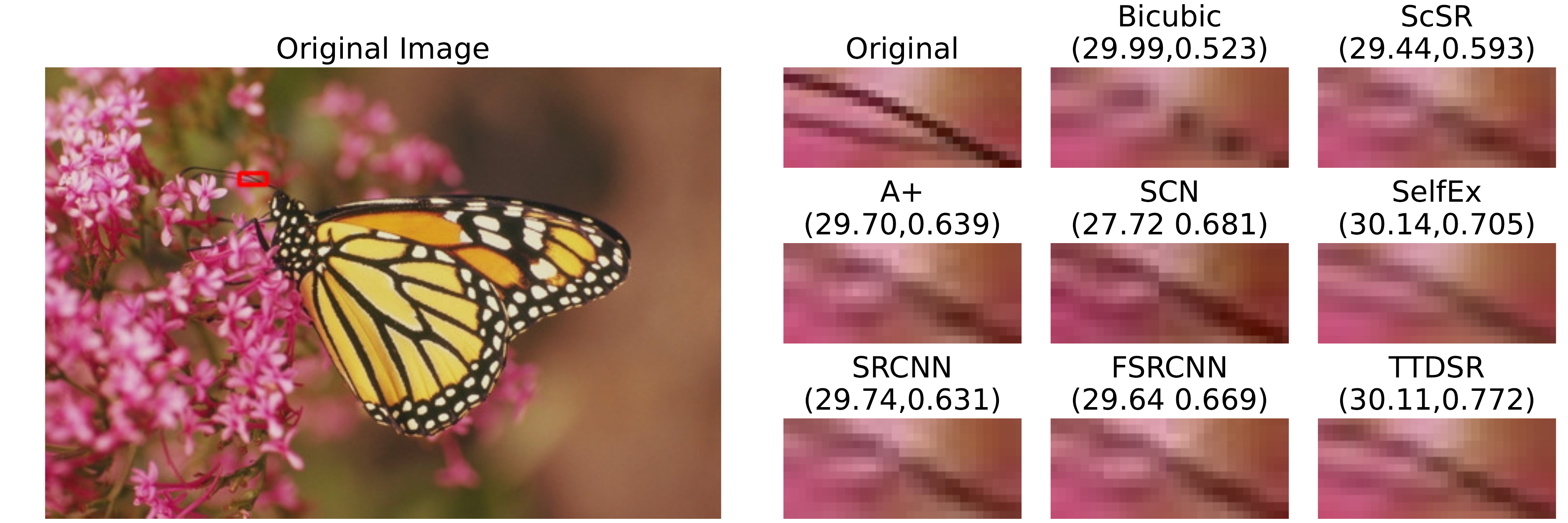} \\
\end{tabular}
\caption{The SR results on image $monarch.bmp$ from Set14 for scale factor of 3 using different methods. Results are represented in the form of (PSNR,SSIM).}
\label{fx8}
\end{figure*}

\begin{figure*}[!htbp]
\centering
\begin{tabular}{cc}
    \includegraphics[width=155mm,height=45mm]{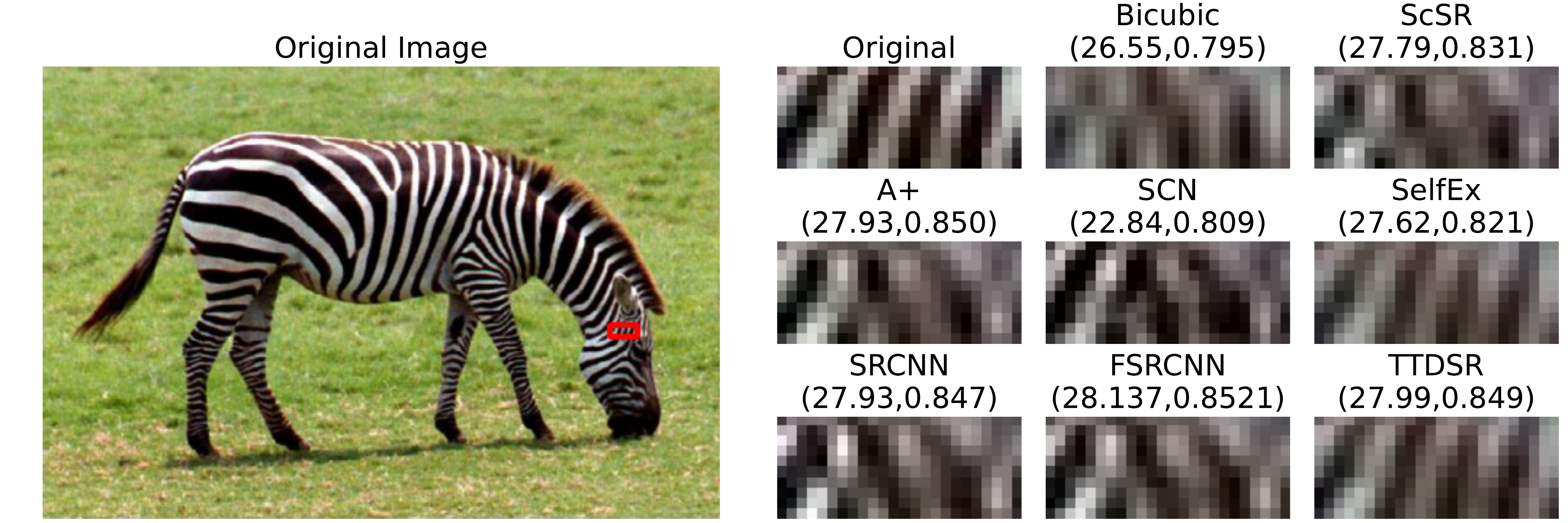} \\
\end{tabular}
\caption{The SR results on image $zebra.bmp$ from Set14 for scale factor of 3 using different methods. Results are represented in the form of (PSNR,SSIM).}
\label{fx7}
\end{figure*}

\begin{table*}[!htbp]
\centering
\renewcommand{\arraystretch}{1.0}
\captionsetup{justification=centering, labelsep=newline,font=footnotesize,labelfont=normalsize}
\caption{SSIM COMPARISON FOR VARIOUS SR METHODS. BLUE  REPRESENTS FIRST AND RED REPRESENTS SECOND BEST.}
\begin{tabular}{cccccccccc}
 \specialrule{1.5pt}{1pt}{1pt} \\
 \multirow{2}{*}{Dataset} & Scale & Bicubic & ScSR & A+ & SelfEx & SCN & SRCNN & FSRCNN &  TTDSR\\
 & & & \cite{9} &\cite{57} &\cite{50} &\cite{58} &\cite{13} &\cite{17} &  (Proposed) \\
    
 \hline \\
 \multirow{3}{*}{Set5} & 2x & 0.9292 & 0.9485 & 0.9544 & 0.9538 & 0.9540 & 0.9542 & \color{red}0.9558 & \color{blue}\textbf{0.9670} \\
    & 3x & 0.8678 & 0.8869 & 0.9088 & 0.9092 & 0.9080 & 0.9090 & \color{red}{0.9140} &  \color{blue}\textbf{0.9154}\\
    & 4x & 0.8101 & 0.8263 & 0.8605 & 0.8640 & 0.8630 & 0.8628 & \color{red}0.8657 & \color{blue}\textbf{0.8774}\\
 \hline \\
  \multirow{3}{*}{Set14} & 2x & 0.8683 & 0.8940 & 0.9055 & 0.9032 & 0.9050 & \color{red}0.9063 & \color{blue}0.9088 &  0.8992\\
    & 3x & 0.7737 & 0.7977 & 0.8188 & 0.8196 & 0.8180 & 0.8209 & \color{red}0.8242 & \color{blue}\textbf{0.8264}\\
    & 4x & 0.7023 & 0.7218 & 0.7489 & 0.7518 & 0.7510 & 0.7503 & \color{red}0.7535 & \color{blue}\textbf{0.7834}\\
 \hline \\
  \multirow{3}{*}{BSDS100} & 2x & 0.8425 & 0.8744 & 0.8864 & 0.8855 & 0.8850 & \color{red}0.8879 & \color{blue}0.8920 & 0.8747\\
    & 3x & 0.7382 & 0.7647 & 0.7836 & 0.7778 & 0.7910 & 0.7863 & \color{red}0.7897 & \color{blue}\textbf{0.7984}\\
    & 4x & 0.6672 & 0.6983 & 0.7087 & 0.7106 & 0.7110 & 0.7101 & \color{red}0.7201 & \color{blue}\textbf{0.7541}\\
 \hline \\
 \multirow{3}{*}{Urban100} & 2x & 0.8408 & 0.8828 & 0.8938 & 0.8967 & \color{red}0.8970 & 0.8946 & \color{blue}0.9010 &  0.8729\\
    & 3x & 0.7349 & 0.7827 & 0.7973 & 0.7864 & \color{blue}0.8016 & \color{red}0.7989 & 0.7512 &  {0.7848}\\
    & 4x & 0.6573 & 0.7024 & 0.7186 & \color{blue}0.7374 & 0.7260 & 0.7221 & 0.7270 &  \color{red}0.7324\\
 \specialrule{1.5pt}{1pt}{1pt} \\
\end{tabular}
\label{tab:2}
\end{table*}
Tables \ref{tab:1} and \ref{tab:2} report the PSNR and SSIM results of TTDSR and other methods, respectively. Out of all the methods, FSRCNN and VBPS gives competitive score 
when compared with TTDSR. However, TTDSR on an average performs well on all kinds of datasets. We now make a subjective comparison of SR results using various methods. 
Figs. \ref{fx7} and \ref{fx8}  show the SR results for the TTDSR and other methods in enhancing the quality of the image degraded due to bi-cubic interpolation. 
Fig. \ref{fx8} shows the SR results on $monarch$ image. The enlarged version shows the thin black edge at head of monarch image. It can be observed that bi-cubic interpolated image shows heavy loss of thin edge details with discontinuity. Also, other methods fails to generate the edges gracefully. The proposed TTDSR architecture generates clear edge and overcomes the discontinuity artifact observed in other methods and gives better results in terms of PSNR and SSIM. Fig. \ref{fx7} shows the SR results on bi-cubic interpolated $zebra$ image. It can be observed that the black and white strips present on $zebra$ lacks details and fails to capture the orientation of the edges. Further, the FSRCNN method shows slightly better results compared to our method in terms of PSNR and SSIM, but the diagonal edges are overlapping leading to poor image visualization. On the other hand, though TTDSR gives second best results but it exploits the frequency domain details to overcome this degradation and thereby, generate correctly oriented black and white strips. 

\begin{figure*}[!htbp]
\centering
\begin{tabular}{cc}
\includegraphics[width=155mm,height=125mm]{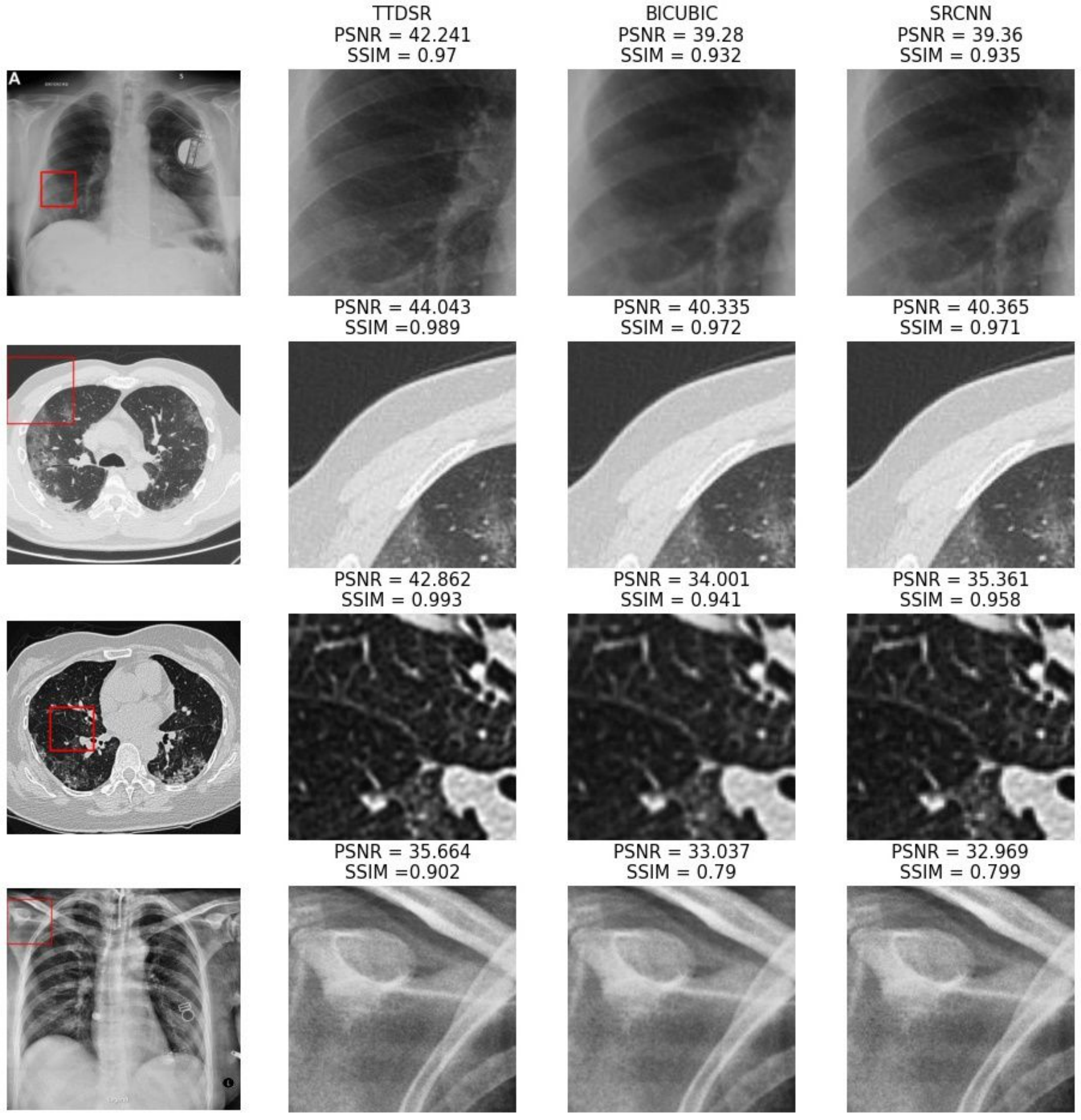} \\
\end{tabular}
\caption{Comparison of SR results on COVID-19 image dataset using different methods. Results are evaluated on scale factor of 2. }
\label{fx11}
\end{figure*}

Next, SR results on COVID-19 medical images is carried out.
Here, the aim is not to detect the infection of COVID-19 through images, but to enhance the quality of these image to provide better diagnosis. The dataset used for this purpose is presented in Cohen $et\;al.$ \cite{54}. As our model is trained on single channel, i.e. Y(luminance) and medical images are gray-scale image that contains only luminance information of the pixel and no color information. This provides a flexibly to use TTDSR architecture on medical images using transfer learning. The SR results on the image can be seen in Fig.\ref{fx11}. The average PSNR and SSIM comparison on COVID-19 dataset can be seen in Table \ref{tab:3}. It can be observed that the proposed method gives better results compared to other methods.

\begin{table}[!htbp]
\centering
\renewcommand{\arraystretch}{1.3}
\captionsetup{justification=centering, labelsep=newline,font=footnotesize,labelfont=normalsize}
\caption{PSNR/SSIM COMPARISON ON COVID-19 DATASET}
\begin{tabular}{ ccccc } 
 \specialrule{1.5pt}{1pt}{1pt} \\
  \multirow{2}{*}{Dataset} & Scale & Bicubic & SRCNN & TTDSR\\
 & & & & (Proposed) \\
  \hline \\
  \multirow{3}{*}{COVID-19} & 2x & 41.32/0.9419 & 41.76/0.9493& \textbf{43.43/0.9806} \\
    & 3x & 40.19/0.9248 &  40.25/0.9281 & \textbf{40.79/0.9681}\\
    & 4x & 39.09/0.9059 &  38.96/0.9077 & \textbf{39.83/0.9351}\\
 \specialrule{1.5pt}{1pt}{1pt} \\
\end{tabular}
\label{tab:3}
\end{table}

\subsection{Network Parameters and its impact}
\subsubsection{Split Point for Tchebichef Frequency Cube}
\begin{figure}[h!]
  \centering
  \includegraphics[width=250pt,height=150pt]{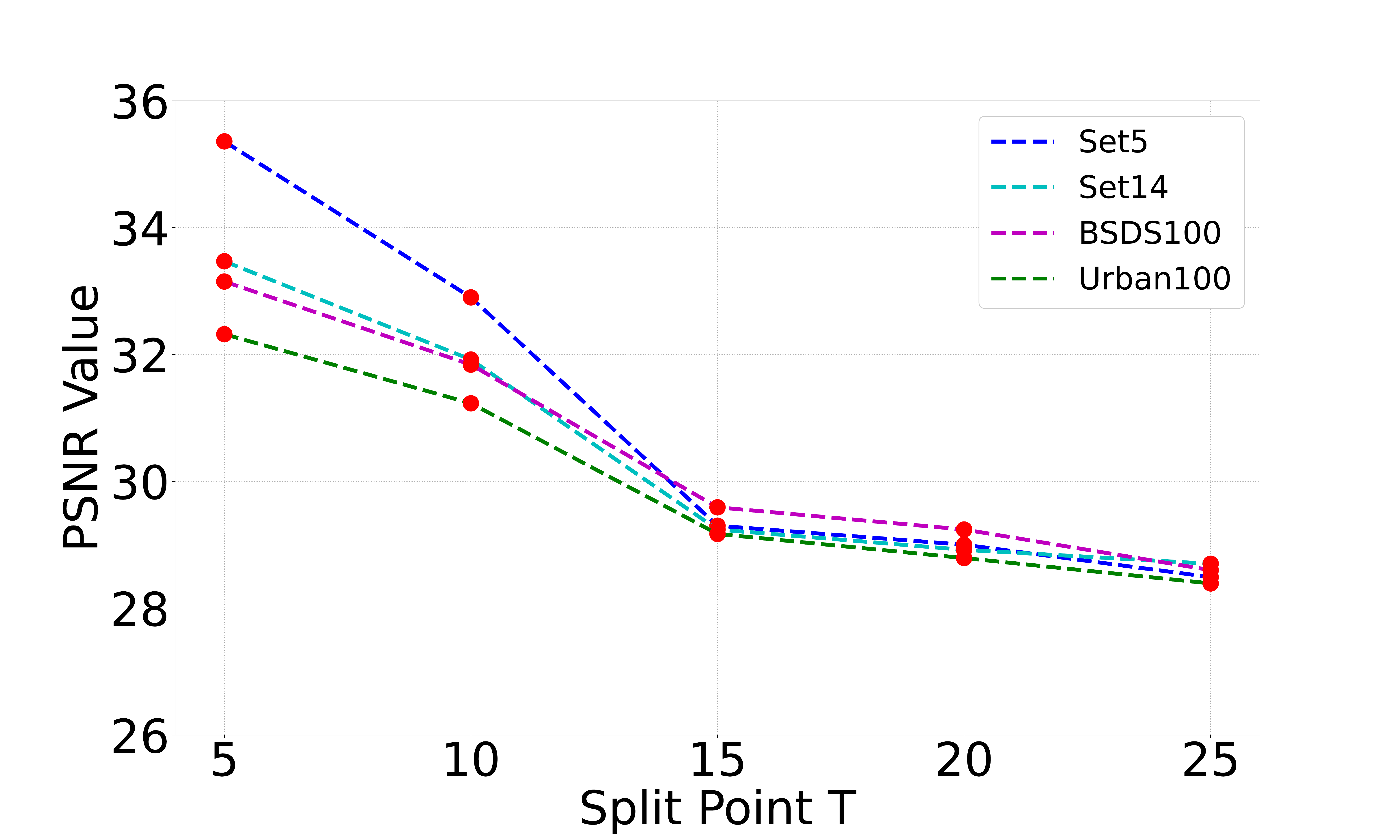}
  \captionof{figure}{ Average PSNR of Set5, Set14, BSDS100 and Urban100 datasets using different values of T.}
\label{fx9}
\end{figure}

Tchebichef polynomials are treated as filters and create a frequency cube in the image transform domain as shown in Fig. \ref{fx5}. In the architecture (Fig. \ref{fx4}), it can be observed that there are two sub-networks, one for recovering the loss in high-frequency content and the other works for recovering loss in low-frequency content. The frequency cube is split into two halves at a split point T. The splitting of the frequency cube is experimental and the performance of the network varies with the different values of T. In this paper, the value of $T$ is taken as 5 based on the experiment conducted in Fig. \ref{fx9}, where for different datasets, the average PSNR reported by the network is highest when T is taken as 5. 

\subsubsection{Impact of Residual Connection}
\begin{figure}[ht!]
  \centering
  \vspace{1ex}%
  \includegraphics[width=250pt,height=200pt]{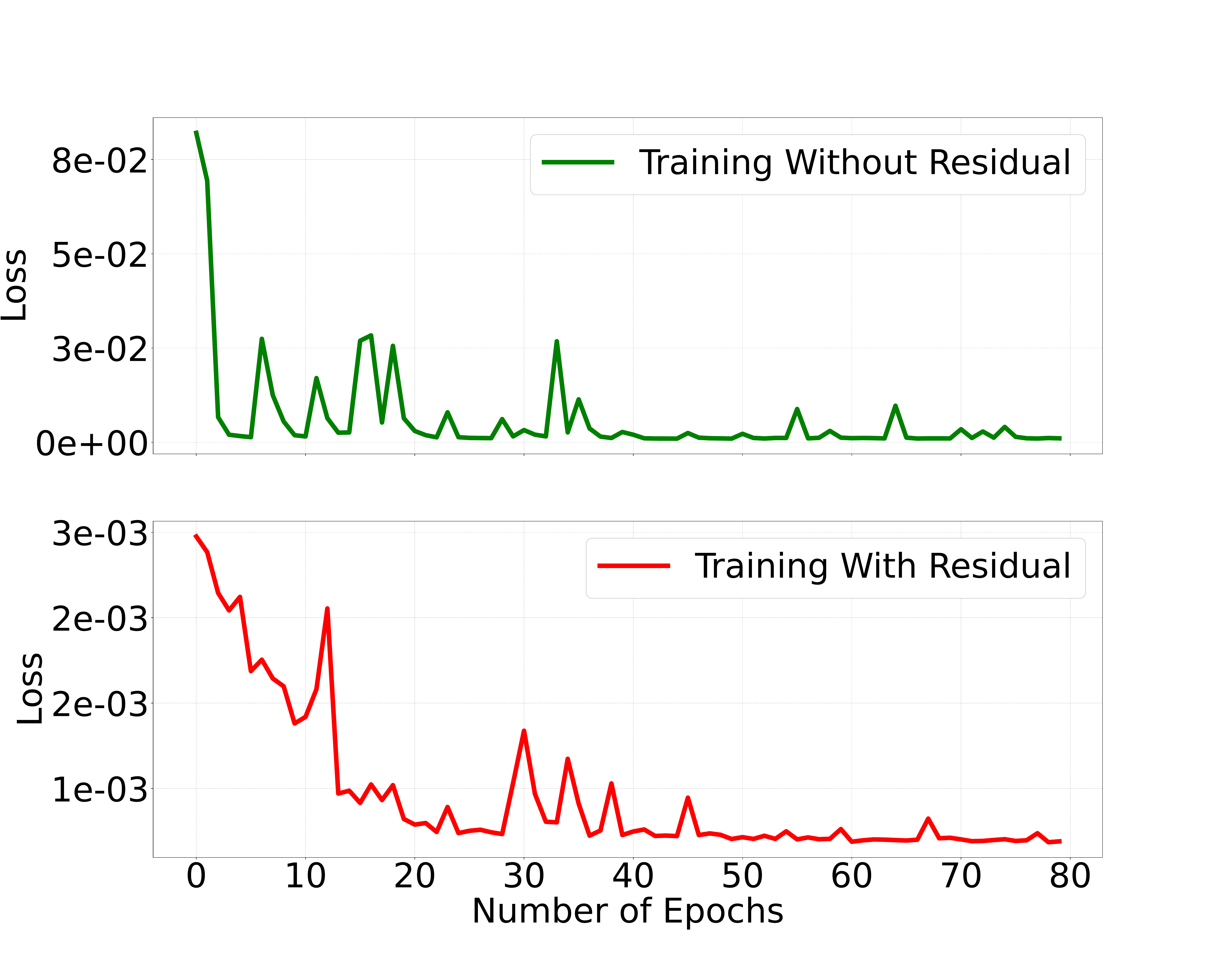}
  \captionof{figure}{Training loss with and without residual architecture.}
  \label{fx10}
\end{figure}

Initially, the network was structured without the residual connection and only relied on the inception based module. In this case the network  performed with limited capability due to vanishing gradient problem. To overcome this problem, two major residual connections are added in the network, local residual connection for high frequency components ($\textbf{f}_{high}$) and the overall global residual connection for both high and low frequency component.
The local residual connection \cite{42} in the inception module was introduced to boost the gradients in the training phase for the recovery of high-frequency components of the image. The use of local residual connectivity overcomes the vanishing gradient problem and helps the optimizer to reach the minima faster.  Experimental analysis for same is shown in Fig. \ref{fx10}, where it can be observed that the architecture with residual connections converges to smaller loss $L$, when used without residual connections. 

\subsubsection{Optimized Learned Tchebichef filters}
\begin{figure}[ht!]
  \centering
  \includegraphics[width=250pt,height=70pt]{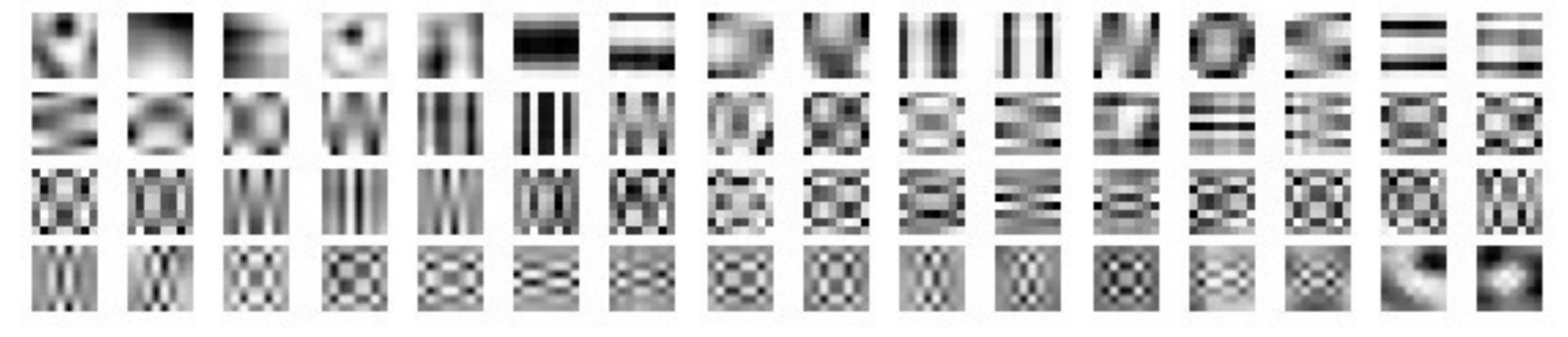}
  \captionof{figure}{Optimized Tchebichef kernels obtained after training process}
\label{fx12}
\end{figure}

As discussed, the network architecture consists of two custom layers namely $TCL$ and $ITCL$. The kernel functions used in $TCL$ are fixed; whereas in that of $ITCL$ are kept trainable to adapt to the training phase of the network. Fig. \ref{fx12} shows the optimized kernels obtained after training process. These optimized kernels are used to reconstruct the image from Tchebichef moment domain and hence gives part of its contribution in providing better image quality when compared to other methods. 

\section{Conclusion}
A deep learning architecture for super resolution of natural and COVID-19 medical images is presented. It makes use of Tchebichef transform domain that helps in exploiting the low and high frequency details present in the images to enhance its quality. The detail analysis of various parameters that affects the performance of the architecture is discussed. The objective and visual comparison of the SR results with the existing methods shows that the proposed architecture provides superior results when evaluated in terms of average PSNR and SSIM metrics. The visual comparison of the result shows that our work restore the details present in an image effectively. This work  opens up the possibility to explore SR architectures using different transform domains.

\bibliographystyle{unsrt}
\bibliography{references.bib}
\end{document}